\documentclass[
reprint,
superscriptaddress,
 nolongbibliography,
 amsmath,amssymb,
 aps,
 prd
]{revtex4-2}

\usepackage{graphicx}
\usepackage{dcolumn}
\usepackage{natbib}
\usepackage{bm}
\usepackage[hidelinks]{hyperref}

\usepackage{braket}
\usepackage{slashed}
\usepackage{upgreek}

\newcommand{\ii}{i}
\renewcommand{\a}{\mathrm a}
\newcommand{\e}{\mathrm e}
\renewcommand{\P}{\mathrm P}
\newcommand{\f}{\mathrm f}
\newcommand{\s}{\mathrm s}
\begin{document}

\preprint{}

\title{Constraints on Axion Dark Matter by Spin-Dependent Macroscopic Force}

\author{Dongyi Yang}
\affiliation{School of Physics, Peking University, Beijing 100871, China}

\author{Chenxi Sun}
\affiliation{State Key Laboratory of Advanced Optical Communication Systems and Networks, School of Electronics,\\
and Center for Quantum Information Technology, Peking University, Beijing 100871, China}

\author{Jianwei Zhang}
\email{james@pku.edu.cn}
\affiliation{School of Physics, Peking University, Beijing 100871, China}






\begin{abstract}

Axion-like particles (ALPs) are hypothetical particles that serve as promising candidates for cold dark matter. Portals like inelastic axion scattering and axion propagated force have been employed to search for the upper limit of the ALPs' coupling with standard model particles. Other methods, like the axion-fermion interaction in the CASPEr experiment, integrate the dark matter motion into the measurement. We suggest a new method for detecting dark matter axions based on axion-electron elastic scattering. In the pseudoscalar axion model, this interaction can be seen as an effective magnetic field, so high-sensitivity atomic magnetometers can be utilized to measure this interaction. The scattering cross section of this process is significantly amplified by the high number density and occupation number of axion dark matter. The upper limit of the electron-axion coupling coefficient obtained from this process can reach two orders of magnitude higher than previous results at low axion mass, and will exceed the astrophysics limits by using a developing magnetometer. This scattering process also provides an efficient way to detect local structures of dark matter.

\end{abstract}

\maketitle

\section{Introduction}
The standard model \cite{[{See, for example }]mann_introduction_2009,*Maiani_Standard_2016} serves as the cornerstone of modern physics, providing an effective lagrangian based on six quarks, six leptons, four gauge bosons, the Higgs boson, and their interactions. Although the standard model has become widely recognized as the most basic particle physics model, there remain intriguing phenomena, such as dark matter \cite{carroll_dark_2007} and the muon $g-2$
experiment \cite{muon_g2_collaboration_measurement_2021,ge_probing_2021,buen-abad_challenges_2021}, that challenge the explanatory power of the standard model, prompting to delve into physics beyond its confines. Among the intriguing possibilities is the introduction of additional particles within the theoretical framework of quantum field theory, offering a promising avenue of exploration without the need for entirely new fundamental physics frameworks.

One compelling candidate in this quest for new particles is the quantum chromodynamics (QCD) axion \cite{weinberg_new_1978,Wilczek_Problem_1978}, a scalar boson that extends beyond the boundaries of the standard model and was initially proposed during investigations into the strong CP (charge conjugation and parity reversal) problem \cite{peccei_mathrmcp_1977}. Axions also feature prominently in discussions about dark matter \cite{nomura_dark_2009,chadha-day_axion_2022}, considered a plausible constituent of this elusive cosmic substance. Axions also play a significant role in addressing various other phenomena, known as axion-like particles (ALPs) which are not required to solve the strong CP problem \cite{chadha-day_axion_2022}. The restriction for QCD axions is so strong that the possibility of QCD axions serving as a component of dark matter is sometimes questioned \cite{Leslie_Dark-matter_2015}. Nevertheless, since ALPs share a broader parameter space compared to the QCD axions, they still serve as a very promising candidate for dark matter, and our discussion in this work focuses on ALPs.

\begin{figure}[t]
    \centering
    \begin{tabular}{c}
        \includegraphics[width = 0.9\columnwidth]{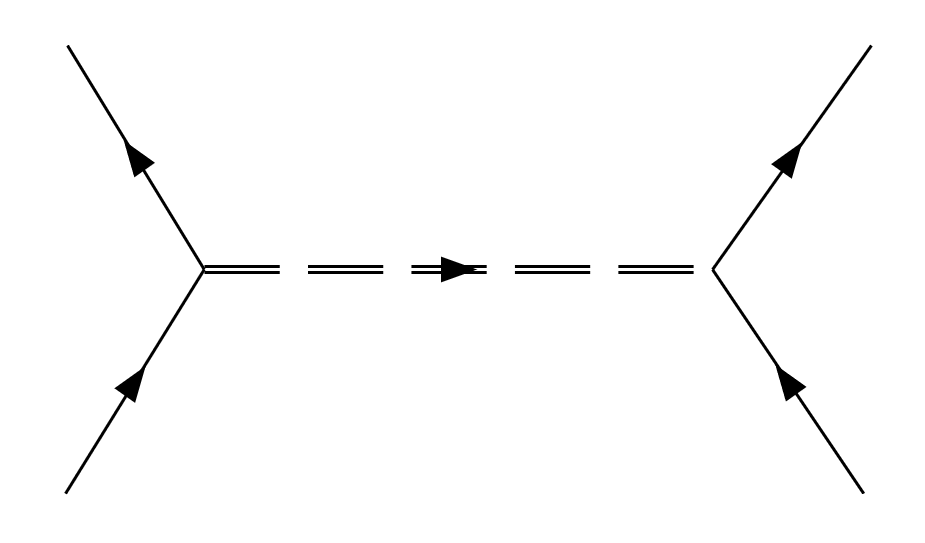} \\
        (a) \\
        \includegraphics[width = 0.9\columnwidth]{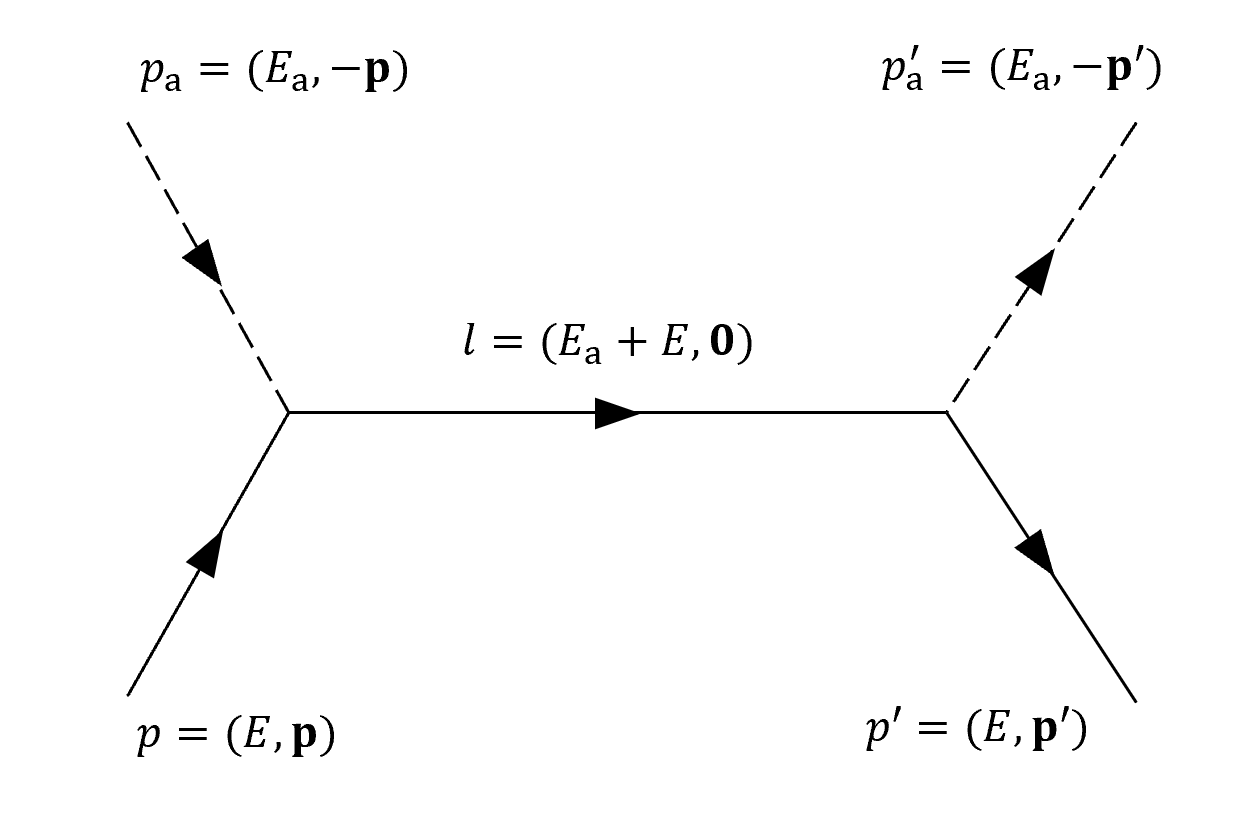}\\
        (b)
    \end{tabular}
    \caption{The Feynman diagram for (a) interaction between two fermions propagated by a hypothetical boson; (b)the scattering process between fermion and axion. Here, the solid lines represent fermions like electrons, neutrons, and protons, the double-dashed line represents the hypothetical bosons such as axions, paraphotons, and Z' bosons, and the dashed lines represent axions specifically.}
    \label{fig:feynman}
\end{figure}

In the exploration of exotic physics, the ansatz that there exists a new macroscopic force apart from the four basic interactions, as known as the fifth force, is also of great concern \cite{moody_new_1984,fischbach_six_1992}. Such macroscopic force describes a weak interaction between standard model fermions like electrons and nucleons propagated by exotic bosons such as ALPs, $\mathrm Z'$ bosons and paraphotons \cite{fadeev_revisiting_2019}. In 2006, a complete set of 16 possible vector forms of this new macroscopic force was determined by Dobrescu and Mocioiu \cite{dobrescu_spin-dependent_2006}. The Feynman diagram of such interactions is demonstrated as \autoref{fig:feynman}(a).

This force should be extraordinarily weak since it is proportional to the square of the coupling coefficient $g_\P$ between the exotic boson and fermion \cite{dobrescu_spin-dependent_2006}. To detect this weak interaction and determine its upper limit, the traditional approach is to use the accelerator to study particle excitations at high-energy levels \cite{K.Barth_2013}. Due to the macroscopic nature of this force, it can be explored at lower energy levels, which gives us a more cost-effective and accessible way to detect this exotic force. Some models have investigated the implications of this exotic interaction on a stellar scale, such as in supernova explosions \cite{bollig_muons_2020}. Other methods involve conducting ultra-high precision experiments in a laboratory setting to measure this weak force. For example, among the 16 non-relativistic interactions predicted in Ref. \cite{dobrescu_spin-dependent_2006}, there exists a specific form of interaction between polarized fermion and unpolarized fermion, 
$
    V_{4+5} \propto \boldsymbol \sigma \cdot (\mathbf v\times \mathbf r),
$
where $\boldsymbol \sigma$ denotes the polarized spin, $\mathbf v$ denotes the relative velocity, and $\mathbf r$ denotes the relative position. This form of interaction is similar to the interaction between spin and a magnetic field, so it can be measured as an effective magnetic field to obtain the upper limit of its coupling coefficient \cite{kim_experimental_2018,ding_constraints_2020,su_search_2021,wu_experimental_2022,ji_constraints_2023,xiao_femtotesla_2023,wei_constraints_2022}. However, this specific form of interaction only exists in the scalar axion model and the $\mathrm Z'$ boson model \cite{dobrescu_spin-dependent_2006,fadeev_revisiting_2019,wei_constraints_2022}. In the pseudoscalar axion model we are discussing, the interaction between fermions does not conform to this form \cite{daido_sign_2017,capolupo_probing_2020}.

A common method in searching for axions is to utilize the scattering process between axions and other particles, such as inelastic axion-fermion scattering that emits photons (also known as the Primakoff effect) \cite{abe_atomic_2021,atlas_collaboration_search_2023,wang_axion-pion_2024}, the axion decay process \cite{dent_new_2020}, and B-meson decays \cite{altmannshofer_b_2017,babar_collaboration_search_2022,lhcb_collaboration_analysis_2022}. Some scattering processes have integrated the dark matter motion into consideration. For example, in the CASPEr experiment \cite{budker_proposal_2014}, the measurement is related to the velocity of dark matter wind. Nevertheless, there are currently few theoretical studies on the elastic scattering process between axions and fermions, such as electrons, as shown in \autoref{fig:feynman}(b). In this article, we propose that the non-relativistic approximation of this interaction has the same form as $V_{4+5}$, which allows us to investigate the idea of using effective magnetic field measurement to detect possible dark matter particles. The intensity of this interaction is influenced by both the electron field and the dark matter axion field. Given that dark matter makes up the majority of the universe's total mass and axions have very small static masses and coupling coefficients, a significant number density of axions is required if dark matter is composed of them to contribute all or part of the dark matter mass. Combined with current estimates of dark matter angular velocity, such an effective magnetic field could possibly be detected by magnetometers. Finally, we discuss how this method helps to understand the stellar-scale mini-structures of dark matter.

\section{Model}
\subsection{Single particle interaction}
In our model, the axions have a pseudoscalar interaction with the electrons, and the axion part of its lagrangian \cite{dobrescu_spin-dependent_2006} reads
\begin{equation}
    \mathcal L = m_\a \partial^\mu a \partial_\mu a
    +g_{\a\e\e} a \bar \psi \gamma^5 \psi,
\end{equation}
where $a$ denotes the axion field, $\psi$ denotes the electron field, $m_\a$ denotes the mass of axion, and $g_{\a\e\e}$ denotes the coupling coefficient between axions and electrons. The conventions of the Dirac matrices and $\gamma^5$ are shown in Appendix A. This model can also be applied to neutrons and muons, which are only different from electrons by mass. We focus on the elastic scattering process of $\text{axion + electron} \rightarrow \text{axion + electron}$. The corresponding tree-order Feynman diagram of this process is shown in \autoref{fig:feynman}(b). The incoming electron's 4-momentum is denoted as $p = (E_\e,\mathbf p)$ and the outgoing 4-momentum as $p' = (E_\e',\mathbf p')$, where $E_\e = \sqrt{\mathbf p^2 + m_e^2}$ is the electron's energy \footnote{In this article, Lorentz 4-vectors are represented by normal font letters, while their corresponding 3-dimensional spatial vectors are presented in boldface.}. The 4-momenta of the incoming and outgoing axion are denoted as $p_\a$ and $p_\a'$ respectively. The scattering amplitude reads (here and in the following article we assume that $\hbar = c = 1$)
\begin{equation}
    \mathcal M(p,p') = \mathrm i S' \mathcal P_\e (l^2) S,
    \label{eq:111}
\end{equation}
where
\begin{equation}
    \mathcal P_\e(l) = \frac{-\mathrm i (- \slashed l + m_\e)}{l^2+m_\e^2 - \mathrm i \epsilon}
\end{equation}
is the propagator of the electron where $\slashed{l} = \gamma^\mu l_\mu$, $l$ is the total 4-momentum and $\ii ^2 = -1$. The node functions $S$ and $S'$ read
\begin{equation}
\begin{aligned}
    S &= (\mathrm i g_{\a\e\e} \gamma^5) u(p),\\
    S' &= \bar u(p')(\mathrm i g_{\a\e\e} \gamma^5).
\end{aligned}
\end{equation}
Here $u(p)$ is the wave function of the incoming electron and $\bar u(p') = u^\dagger (p') \gamma^0$ is the wave function of the outgoing electron. To simplify our analysis, we will calculate the scattering amplitude in the center of mass (CM) frame. We use a different set of 4-momentum variables, denoted as $q = (q_0,\mathbf q)$ and $P = (P_0,\mathbf P)$, where $q = p - p'$ represents the 4-momentum transferred during the interaction, and $P = (p + p')/2$ denotes the average 4-momentum of the electron. Energy-momentum conservation implies that $|\mathbf p| = |\mathbf p'|$, or equivalently $q_0 = E_\e -E_\e' = 0$, which is to say that no energy is transferred during the elastic scattering. In the CM frame, we have (see Appendix B for detailed derivation of these two formulas)
\begin{equation}
    \bar u(p') \gamma^0 u(p) = E_\e+m_\e + \frac{1}{E_\e+m_\e}\left(
\mathbf P^2-\mathbf q^2/4 + \mathrm i (\mathbf q \times \mathbf P)\cdot \boldsymbol \sigma \right),
\label{eq:sppr1}
\end{equation}
\begin{equation}
    \bar u(p') u(p) = 2\sqrt{E_\e^2-\mathbf P^2}\left(
1-\frac{\mathrm i }{2(E_\e^2-\mathbf P^2)}(\mathbf q\times \mathbf P)\cdot\boldsymbol \sigma
\right),
\label{eq:sppr2}
\end{equation}
where $\boldsymbol\sigma$ is the spin polarization of the electron whose norm equals $1$ when the electron is fully polarized. Substitute them into Eq.(\ref{eq:111}) gives
\begin{widetext}
    \begin{equation}
        \mathcal M(p,p') = \frac{g_P^2}{l_0^2 - m_\e^2}
        \left[l_0 E_\e + l_0 m_\e +2m_\e\sqrt{E_\e^2-\mathbf P^2} + \frac{l_0}{E_\e+m_\e}(\mathbf P^2-\mathbf q^2/4)
        + \ii \left(\frac{l_0}{E_\e+m_\e} - \frac{m_\e}{\sqrt{E_\e^2 - \mathbf P^2}}\right) 
        (\mathbf p\times \mathbf p')\cdot\boldsymbol \sigma\right].
\label{eq:scattering_amplitude_explicit}
\end{equation}
\end{widetext}
The approximation here is that we only consider the tree diagram. Since higher-order diagrams are suppressed by orders of $g_{\a\e\e} < 10^{-10}$, the tree-level diagram is sufficient for our discussion.

The electron mass is well known to be around $0.511\ \mathrm{MeV}$ and axion mass is estimated to be of the order $\mathrm{\upmu eV}$ to $\mathrm{eV}$ \cite{peccei_strong_2008}. Although the stationary mass of the electron is much larger than the axion, the kinetic energy is predominantly contributed by the axion since we are at the CM frame. We can make the approximation that
\begin{equation}
    m_\e \gg m_\a \gg \mathbf p^2/m_\a  \gg \mathbf p^2/m_\e ,
\end{equation}
and the total energy is approximately
\begin{equation}
    l_0 = E_\e + E_\a \approx  m_\e+m_\a + \frac{p^2}{2 m_\a}.
\end{equation}
Also, considering that the exchanged momentum $\mathbf q$ is small due to a small coupling constant and non-relativistic approximation, we have
\begin{equation}
    \mathcal M(\mathbf P,\mathbf q) \approx \frac{g_{\a\e\e}^2 m_\a m_\e}{2m_\a^2 + \mathbf P^2 + \frac{1}{4} \mathbf q^2} \left(
    4 + \ii \frac{1}{2m_\e^2} (\mathbf P \times \mathbf q) \cdot \boldsymbol \sigma
    \right).
    \label{eq:amplitude}
\end{equation}

We can determine the semi-classical potential of this interaction by Born approximation. Born approximation states that the semi-classical potential is proportional to the Fourier transform of the corresponding scattering amplitude about the exchanged momentum $\mathbf q$ while keeping the average momentum $\mathbf P$ to be a constant. As a brief proof of this statement, we remind that if the system undergoes a potential $\hat V$, its scattering amplitude from a free particle state with momentum $p$ into another free particle state with momentum $p'$ can be written as 
\begin{equation}
    \mathcal M (p,p') \propto \bra{p'} \hat V \ket{p}.
\end{equation}
In our case, the potential should be of the form $V(\mathbf r,\mathbf v) = \boldsymbol \sigma \cdot (\mathbf p \times \mathbf r ) f(r)$ classically. When quantizing this potential, noting that the momentum and position operators do not commute, its operator form should be
\begin{equation}
    \hat V = \frac{1}{2}\hat{\boldsymbol \sigma} \cdot (\hat{\mathbf p} \times \hat{\mathbf r} - \hat{\mathbf r}\times \hat{\mathbf p}) f(r).
\end{equation}
Sandwiched by the free particle states, the Born approximation gives
\begin{equation}
    \mathcal M (p,p') \propto \bra{p'}\hat{\boldsymbol \sigma} \cdot (\mathbf P \times \hat{\mathbf r}) f(r)\ket{p},
\end{equation}
which implies that we should keep the average momentum $\mathbf P$ as a constant and only perform the Fourier transform on the exchanged momentum $\mathbf q$. As a result, the spin- and velocity-dependent semi-classical potential of axion-fermion elastic scattering is
\begin{widetext}
    \begin{equation}
    V(\mathbf r,\mathbf v) = \frac{1}{4E_\a E_\e}\int \mathrm d \mathbf q^3 \e^{\ii \mathbf q\cdot \mathbf r} \mathcal M(\mathbf P,\mathbf q)
     = -2g_{\a\e\e}^2 \frac{1}{8\pi m_\e}
    [\boldsymbol{\sigma} \cdot(\mathbf{v} \times \hat{\mathbf{r}})]
    \left(\frac{1}{r^2}+\frac{1}{\lambda r}\right) \e^{-r/\lambda},
    \label{eq:11111}
\end{equation}
\end{widetext}
where $\lambda = 1/(2\sqrt{2} m_\a)$ is the force range, $\mathbf v$ is the relative velocity of the electron and axion, $\mathbf r$ is the relative position, and $\hat{\mathbf r} = \mathbf r / |\mathbf r|$ is the unit vector of $\mathbf r$. This fermion-axion interaction has an almost identical form to the fermion-fermion interaction $V_{4+5}$ in Ref. \cite{dobrescu_spin-dependent_2006}. The only two differences are that there is an additional factor of 2 due to the symmetry of the Feynman diagram, and the force range $\lambda$ here has an extra factor $1/{2\sqrt{2}}$ compared to the force range of $V_{4+5}$.

\subsection{Enhancement effects}
Eq.\eqref{eq:11111} shows how a single electron interacts with a single axion. In reality, we need to consider how a group of electrons interacts with a group of axions. Specifically, in magnetometers, the effective magnetic field is related to the effective interaction between a single electron and the dark matter axions. Since the axion has a small mass $m_\a$, it processes a macroscopic force range. As a result, a single electron will interact with a large number of axions within this force range simultaneously, generating an enhancement proportional to the dark matter axion density $n_\a$. Since the dark matter possesses an energy density $\rho_{\mathrm{DM}} = 0.3 \pm 0.1 \ \mathrm{GeV/cm^3}$ \cite{sofue_rotation_2020}, it gives a enormous axion number density $n_\a = \rho_{\mathrm{DM}}/m_\a \approx 0.3\times 10^{15} \ \mathrm{cm^{-3}}\ \left(1\ \mathrm{\upmu eV}/m_\a\right)$ \cite{peccei_strong_2008}.

Another enhancement effect is called the stimulated emission enhancement. Since axions are bosons, the scattering cross section will be multiplied if there's a degeneracy of the outgoing state $\ket{\a_{\mathrm{out}}}$ \cite{fukuda_detection_2022}. This enhancement occurs from the exchange symmetry of identical bosons. The scattering amplitude $\mathcal M$ we calculated previously is the transition amplitude from the incoming single-particle state to the outgoing single-particle state, i.e.
\begin{equation}
    \mathcal M = \bra{\e_{\text{out}}}
    \bra{\a_{\text{out}}} \hat S \ket{\a_\text{in}}
    \ket{\e_{\text{in}}},
\end{equation}
where $\hat S$ is the action and $\ket{\e/\a_{\mathrm{in/out}}}$ denotes the incoming/outgoing state of the electron/axion. In reality we are interested in the transition amplitude from a many-particle state to another many-particle state which differs from the single-particle case. A many-particle state of an identical boson can be decomposed into
\begin{equation}
    \ket{\text{in/out:}\mathcal{N}}\otimes\ket{\text{BG}} = 
    \frac{1}{\sqrt{\mathcal{N} !}}  (\hat a^\dagger_{\text{in/out}})^\mathcal{N}
    \ket{\text{in/out}:0}\otimes\ket{\text{BG}},
\end{equation}
where $\ket{\text{in/out:}\mathcal{N}}$ denotes the $\mathcal{N}$-degenerate incoming/outgoing state, $\ket{\text{BG}}$ denotes the background state which is orthogonal to the incoming and outgoing state, and $\hat a^\dagger_{\text{in/out}}$ is the creation operator of the incoming/outgoing state. Thus, the many-particle scattering amplitude $\mathcal M_f$ reads
\begin{equation}
\begin{aligned}
    \mathcal M_f =& \bra{e_{\text{out}}}
    \bra{\text{out:}\mathcal{N}'+1 ,\text{in:} \mathcal{N}-1} \hat S \ket{\text{out:}\mathcal{N}' ,\text{in:} \mathcal{N}}
    \ket{e_{\text{in}}}\\
    =&\braket{e_{\text{out}}|e_{\text{in}}}
    \sqrt{\mathcal{N}'+1}
    \bra{\text{out:}1,\text{in:}\mathcal{N}-1}\hat S\ket{\text{out:}0,\text{in:}\mathcal{N}},\\
\end{aligned}
\label{eq:mmanybody}
\end{equation}
where $\mathcal{N}'$ is the degeneracy of the outgoing state and $\mathcal{N}$ is the degeneracy of the incoming state. So, there is an extra factor $\sqrt{\mathcal{N}'}$ which is called the stimulated emission effect.

Although this enhancement effect does exist, it will be canceled out in the scalar part of the scattering amplitudes. The effective force between axions and electrons can be written as \cite{fukuda_detection_2022}
\begin{equation}
    \mathbf F = \int \mathrm d \mathbf p_\ii \int \mathrm d \mathbf p_\f \delta(|\mathbf p_\ii| - |\mathbf p_\f|) \frac{|\mathbf p_\ii|}{m_\a}
    (\mathbf p_\ii - \mathbf p_\f) 
    f_{\mathrm{in}} (1 + f_\mathrm{out}) \frac{\mathrm d \sigma}{\mathrm d \Omega},
\end{equation}
where $\mathbf p_\ii,\ \mathbf p_\f$ are the incoming and outgoing momentum respectively. Since the $f_\mathrm{in} f_{\mathrm{out}} (\mathbf p_\ii - \mathbf p_\f)$ term changes sign after exchanging the incoming state and outgoing state, it becomes zero after integration. So, in the scalar part of this scattering process, the stimulated emission enhancement does not affect the effective force, as concluded by Fukuda and Shirai in Ref. \cite{fukuda_detection_2022}. In our model, this stimulated emission enhancement survives under integration. The spin- and velocity-dependent potential, $V \propto \boldsymbol \sigma \cdot (\mathbf v \times \mathbf r)$, represents the exchange between angular momentum $\mathbf L = \mathbf v\times \mathbf r$ and electron spin $\boldsymbol \sigma$, while not the momentum when we are interested in force. As a result, the exchanged angular momentum is even under exchanging incoming and outgoing states. Thus, such cancelation in the scalar case doesn't hold here anymore, and the spin-dependent potential benefits from stimulated emission enhancement.

Denoting the occupation number of the outgoing state as $\mathcal{\mathcal{N}}$, the enhanced scattering amplitude becomes $\mathcal M = \sqrt{1+\mathcal{\mathcal{N}}} \mathcal M_0$ where $\mathcal M_0$ is the single particle scattering amplitude. Combining these two enhancement effects and assuming that the dark matter axion is spatially uniform and has an angular velocity $\boldsymbol\Omega$ with respect to the inertial frame along the direction of electron spin $\boldsymbol \sigma$, the effective magnetic field is 
\begin{equation}
    B_{\mathrm{eff}}
    = (g_{\a\e\e}^2 \Omega)
    \frac{\rho_{\mathrm{DM}}}{4\gamma m_\e m_\a^3}
    \sqrt{\mathcal{\mathcal{N}}},
    \label{eq:ssssss}
\end{equation}
where $\gamma$ is the gyromagnetic ratio of the electron.

Finally, we analyze the applicable range of our perturbation method. A higher-order scattering process asks for an extra incoming or outgoing axion, which includes an extra incoming or outgoing axion line, an extra node, and an extra electron propagator. The extra axion line will also be enhanced by the density or occupation number. As a result, the relevant dimensionless perturbation factor should be $g_{\a\e\e} \sqrt{\mathcal N}$ on the outgoing side, and $g_{\a\e\e}n_\a V \approx g_{\a\e\e}\rho_{DM}m_\a^{-4}$ on the incoming side. Additionally, the extra propagator contributes a factor
\begin{equation}
    -\frac{\slashed l' - m_\e}{l'^2 + m_\e^2},
\end{equation}
where $l' = l - k_\a'$ and $k_\a'$ is the momentum of the extra axion. The denominator reads
\begin{equation}
    (l_0^2 - m_\e^2) - 2 l_0 E_\a' + m_\a^2 
\approx \frac{m_\e}{m_\a} (\mathbf p^2 - \mathbf k_\a'^2)\sim \frac{m_\e}{m_\a} \mathbf p_\a^2,
\end{equation}
and the numerator reads
\begin{equation}
    \bar u (\slashed l' - m_\e) = \bar u \slashed p_\a \sim p_\a.
\end{equation}
The factor $p_a$ is canceled out through phase space summation, contributing an overall factor $m_a/m_e$.  The mass dependence of the perturbation factors thus reads
\begin{equation}
    g_{\a\e\e}\sqrt{\mathcal N}\frac{m_\a}{m_\e} \approx 10^{3} \left(\frac{1\ \mathrm{\upmu eV}}{m_\a}\right) g_{\a\e\e},
\end{equation}
and
\begin{equation}
    g_{\a\e\e} \rho_{DM} m_\a^{-3}m_\e^{-1} \approx 10^{6} \left(\frac{1\ \mathrm{\upmu eV}}{m_\a}\right)^3 g_{\a\e\e}.
\end{equation}
Therefore, the perturbation theory is valid when
\begin{equation}
    g_{\a\e\e} < 10^{-3} \left(\frac{m_\a}{1\ \mathrm{\upmu eV}}\right),\quad g_{\a\e\e} < 10^{-6} \left(\frac{m_\a}{1\ \mathrm{\upmu eV}}\right)^{3}.
    \label{eq:validity}
\end{equation}

\section{Discussions and Applications}
\subsection{Occupation number}
The occupation number of the axion dark matter gives rise to an enormous amplification of the effective magnetic field. The elastic scattering process is second-order suppressed by the coupling coefficient $g_{\a\e\e}$, which seems to be more unrealistic than those processes with only first-order suppression. However, the inclusion of the outgoing axion turns on the stimulated emission enhancement effect, which might compensate for the flaw included by the second-order coupling coefficient. Therefore, the elastic scattering process is highly sensitive to the phase space distribution of the dark matter axion.

The dark matter axion has a typical velocity, denoted as $v_\s$, which describes its dispersion in the phase space. Given the number density as $n_\a = \rho_\a/m_\a$, assuming that each quantum state occupies $h^3$ volume in the phase space where $h$ is the Planck constant, the occupation number of the dark matter axion reads \cite{fukuda_detection_2022}
\begin{equation}
    \mathcal N \approx \frac{(2\pi)^3 n_\a}{(m_\a v_\s)^3}.
\end{equation}
The typical velocity of dark matter axion in the solar system is estimated as around its circular velocity, approximately $v_\s\approx 250\ \mathrm{km/s}$ \cite{sofue_rotation_2020}. The occupation number of dark matter axion is hence approximately $\mathcal{N} \sim 10^{30} (1\ \mathrm{\upmu eV}/m_\a)^4$ \cite{fukuda_detection_2022}. The incredible occupation number is more than sufficient to compensate for the suppression given by the coupling coefficient.

Some also proposed that axion dark matter forms a Bose-Einstein condensation (BEC) due to its large number density \cite{sikivie_bose-einstein_2009,berges_far_2015}, and hence gives rise to a larger occupation number. If the self-interaction and gravitational interaction are strong enough to rethermalize the dark matter axion, it evolves adiabaticly, and its occupation number is conserved during evolution. Thus, a considerable number of axions stay in the same quantum state, and this BEC state holds a gigantic occupancy. Its typical velocity is of the order of the cosmological expansion, i.e. $v\sim H$ where $H$ is the Hubble coefficient. Thus the occupation number is approximately $\mathcal N_{\mathrm{BEC}}\sim 10^{61} \left(1\ \mathrm{\upmu eV}/m_\a c^2\right)^{8/3}$ \cite{sikivie_bose-einstein_2009,erken_cosmic_2012}. Under this estimation, there will be $16$ orders of amplification to the effective magnetic field. It should be noted that this theory also undergoes some criticisms. It was argued that the self-interaction and gravitational force are not sufficient to rethermalize the axion BEC into its equilibrium state. Besides, the dominant interactions between axions are attractive but not repulsive, which is different from laboratory BEC and prohibits the emergence of long-range correlations \cite{guth_dark_2015,Davidson_axions_2015,Marsh_axion_2016}.

\subsection{Coupling coefficient}
The effective magnetic field generated by elastic scattering naturally exists without any specific conditions and applies to all electrons. It is important to note that this represents an effective precession of the electron spin, rather than a true magnetic field. Consequently, there will be no electric field under Lorentz transformation, and it does not conform to Maxwell's equations. However, since it causes spin precession in a manner similar to a real magnetic field, it can be measured using a spin-based magnetometer. An atomic magnetometer operates by detecting the electron spin precession in response to an external magnetic field, allowing for the identification of potential signals from such scatterings. This device polarizes alkali-metal vapors within a small cell and employs probing light to measure the precession frequency of the electron spin. Additionally, some alternative devices utilize nuclear spin, which can also be incorporated into our model, as neutrons experience a similar interaction with axions \cite{[{For technical details about atomic magnetometers, see for example }]budker_optical_2013}.

The measurement of such a ``background magnetic field'' generated by the elastic scattering can be achieved in a magnetically shielded environment. Such shielding is necessary for some of the atomic magnetometers, such as spin-exchange-relaxation-free magnetometers (SERF), which have to operate under a nearly zero magnetic field. The shielding is also critical to remove noises generated from surroundings. A cylinder barrel made of multilayer high-permeability material is an ideal choice for passive shielding, as commonly used in magnetometer experiments. Magnetic fields generated by sources out of the barrel can be shielded, while the axion effective field will not be affected since it is not a real magnetic field. Besides, the magnetically shielded cylinder itself will not generate any false signals since the axion-propagated force mandates a non-zero rotation \cite{dobrescu_spin-dependent_2006}. Moreover, since the force range is typically within a meter, the magnetically shielded cylinder can be built larger than this size. This ensures that any potential unknown effects are avoided, as the effective magnetic field will be generated entirely within the cylinder. Another possible scheme utilizes active shielding, such as a 3-dimensional Helmholtz coil. It generates a magnetic field to cancel out the noises through feedback from a conventional magnetometer, such as an induction coil or giant magnetoresistance (GMR). Since the scattering process only affects polarized spin, any other measurement of the magnetic field that doesn't involve spin precession will not be influenced by such effects. It also offers a practical method for determining if a non-zero background field results from axion scattering or other unshielded signals.

\begin{figure}[h]
    \centering
    \includegraphics[width=\linewidth]{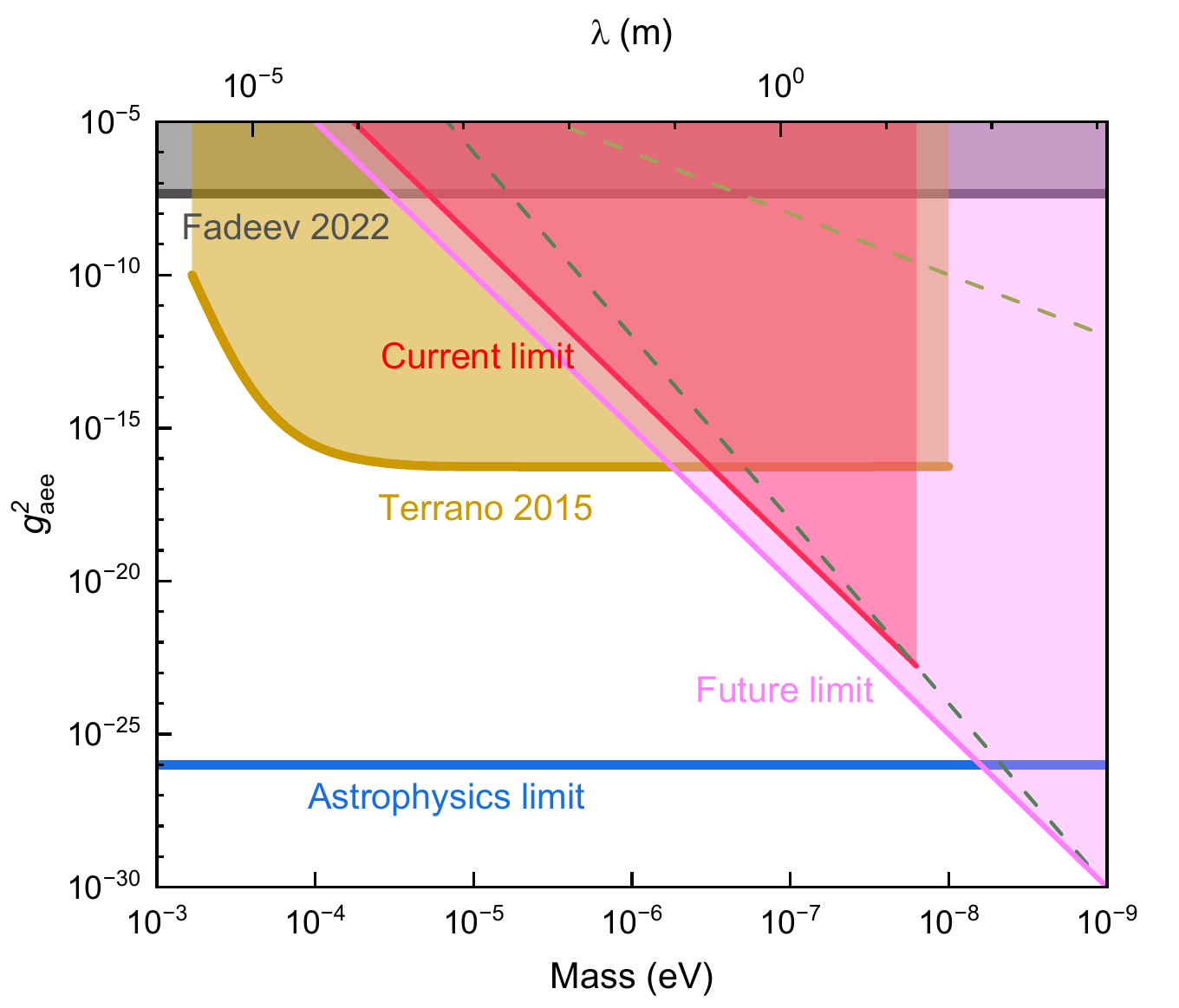}
    \caption{Constraints on axion-electron coupling coefficient as a function of axion mass. The gray line is the upper limit established by Ref. \cite{fadeev_pseudovector_2022}, the yellow line by Ref. \cite{terrano_short-range_2015}, the blue line by Ref. \cite{bertolami_revisiting_2014}. The red line is the upper limit established by this work using the present magnetometer data, and the pink line is established according to a proposed magnetometer with a sensitivity of $10\ \mathrm{aT/Hz^{1/2}}$ \cite{kominis_subfemtotesla_2003}. The dashed lines are the limits for the perturbation theory to be valid, where the lower line is the criterion from dark matter density, and the upper line is from the occupation number.}
    \label{fig:gmplot}
\end{figure}

As we have not measured any of such ``background magnetic fields'', the sensitivity of magnetic field measurements can determine the upper limit of the axion-electron coupling constant and the upper limit of the dark matter angular velocity. Similarly, the upper limit of the axion-nucleon coupling constant can also be determined through measurements using a magnetometer based on nucleon spin. Presently, the highest-sensitivity measurement for magnetic fields reaches $S_B = 0.16 \ \mathrm{fT/Hz^{1/2}}$ \cite{dang_ultrahigh_2010}. In an experiment with time duration $\tau$, the effective magnetic field sensitivity becomes $\Delta B = S_B/\tau^{1/2}$, which in a 270-hour experiment is $\Delta B \approx 0.16 \ \mathrm{aT}$. This provides an upper limit for the coupling coefficient and dark matter angular velocity
\begin{equation}
    g_{\a\e\e}^2 \Omega \lesssim 10^{-29} \ \mathrm{s^{-1}}\left(\frac{m_\a}{1\ \mathrm{\upmu eV}}\right)^{5},
    \label{eq:result}
\end{equation}
where the occupation number is conservatively estimated as $\mathcal N \sim 10^{30} (1\ \mathrm{\upmu eV}/m_\a)^4$. If we apply the BEC model, the upper limit will be further reduced by 15 orders of magnitude. Since the occupation number is roughly estimated, we use the symbol $\sim$ to denote that it is just an approximated value. The dark matter motion is reflected in the angular velocity $\Omega$. If the dark matter itself processes a large enough angular velocity, this potential will generate a detectable ``background magnetic field'' to the atomic magnetometers. The dark halo model and cosmological observations present evidence that a non-zero angular velocity of the order $\Omega_{\mathrm{DH}} \sim 20 \ \mathrm{km/(kpc\cdot s)} \sim 10^{-15} \ \mathrm{s^{-1}}$ \cite{sofue_rotation_2020,obreja_first_2022} exists, where $\mathrm{kpc}$ denotes kilo-parsec. This angular velocity is of the same order as that of the Sun orbiting the center of the galaxy. An upper bound on axion-electron coupling can therefore be established based on our theory, as shown in Fig. \ref{fig:gmplot}. Note that our theory is valid when Eq.(\ref{eq:validity}) is satisfied, which sets a mass region as shown in the figure. Based on the current magnetometer sensitivity, our theory is valid when $m_\a > 0.016 \ \mathrm{\upmu eV}$, with an upper bound on the coupling coefficient of about $10^{-12}$ at the low mass region, which is three orders of magnitude higher than previous results \cite{terrano_short-range_2015}. If we can further upgrade the sensitivity to $10 \ \mathrm{aT/Hz^{1/2}}$, which was proposed theoretically by Ref. \cite{kominis_subfemtotesla_2003} and is being worked on by experimental teams, our theory would be valid when $m_\a> 1\ \mathrm{neV}$, with an upper bound of about $10^{-15}$, which exceeds the astrophysics limits in low mass region \cite{bertolami_revisiting_2014}.

In terms of measuring the upper limit of the coupling coefficient, the fermion-axion interaction has higher sensitivity at low axion mass, but the effect is relatively poor at large axion mass. Although dark matter accounts for most of the mass of matter in the universe, its local number density is still much lower than that of solids since it is distributed almost uniformly in the universe. However, the interaction between fermions and axions is also enhanced by the degeneracy of axions due to the stimulated emission effect, in addition to the enhancement caused by the macroscopic force range. As the number density, degeneracy, and force range of axions increase while decreasing axion mass, the fermion-axion interaction has a larger dependency on axion mass. For comparison, in the case of fermion-fermion interaction, only the force range is affected by the axion mass. Additionally, the rotating matter ring has a fixed radius, which prohibits higher sensitivity when the force range is larger than its radius. Specifically, since the matter ring has to rotate stably without perturbing the magnetometer, in a typical setup, the matter ring has a radius of around several centimeters. As a result, the fermion-axion interaction can give a much better prediction for low-mass axions whose force range is at the order of kilometers.

\subsection{Substructures of dark matter}

The scattering process discussed in this article is similar to the magnetometer experiments \cite{xiao_femtotesla_2023} in terms of single-particle interaction forces. The critical difference is that the force here occurs between fermions and axions rather than between fermions and fermions. In experiments involving fermion-axion interactions, only electrons in magnetometers can be accurately manipulated and measured, unlike in experiments involving fermion-fermion interactions, where both sides of fermions (electrons in magnetometers and nucleons in crystals) can be manipulated and measured accurately. The properties of dark matter axions can only be estimated through theoretical models and other observations. Therefore, the upper limit of the coupling coefficient obtained based on this interaction appears together with the motion of dark matter itself. Although it becomes difficult to obtain a convincing upper limit on the coupling coefficient of the axion through such experiments, our model, when combined with other experiments on the upper limit of the coupling coefficient, can reflect the motion characteristics of the dark matter itself.

For the purpose of discussing the boundaries our model presents for dark matter motion, we fix the coupling coefficient to be $g_{\a\e\e}^2 = 10^{-23}$ \cite{xiao_femtotesla_2023}. Thus an angular velocity larger than $\Omega_{\mathrm{min}} = 10^{-6} \ \mathrm{s^{-1}} \left(m_\a/1\ \mathrm{\upmu eV}\right)^5$ can be detected.  If the occupation number predicted by the BEC theory \cite{sikivie_bose-einstein_2009,berges_far_2015,erken_cosmic_2012} is correct, the minimum angular velocity will be narrowed down to $\Omega_{\mathrm{min}}' = 10^{-21} \ \mathrm{s^{-1}} \left(m_\a/1\ \mathrm{\upmu eV}\right)^5$. The angular velocity of the axion dark matter can originate from various sources. Firstly, dark matter halo gravitationally bounded by a galaxy will circular around the galaxy's core as other celestial bodies do. Cosmological observations have proposed a dark halo spin around $\Omega_{\mathrm{DH}}\sim 200\ \mathrm{km\cdot s^{-1}/10\ kpc}$ \cite{sofue_rotation_2020,obreja_first_2022}, which is of a similar order as that of our Sun. The generation of such an angular velocity is also confirmed by theoretical models \cite{erken_cosmic_2012}. When dark matter axions fall into a galaxy's gravitation well, they will not stay in the lowest energy state due to the time-varying background. The gravitational force between axions is large enough to rethermalize them into a new state consistent with the total angular momentum acquired from neighboring inhomogeneities through tidal torquing \cite{peebles_origin_1969}. Therefore the BEC dark matter model is consistent with the non-zero angular velocity of dark matter, and its angular velocity should have been detected for an axion mass less than about $1\ \mathrm{meV}$. Due to the fact that we did not observe such a signal, it posed difficulties for these models that predicted such a large occupancy number. Since we have not observed such a signal, it poses challenges for the models that predicted such a high occupancy number.

\begin{figure}[t]
    \centering
    \includegraphics[width=0.9\linewidth]{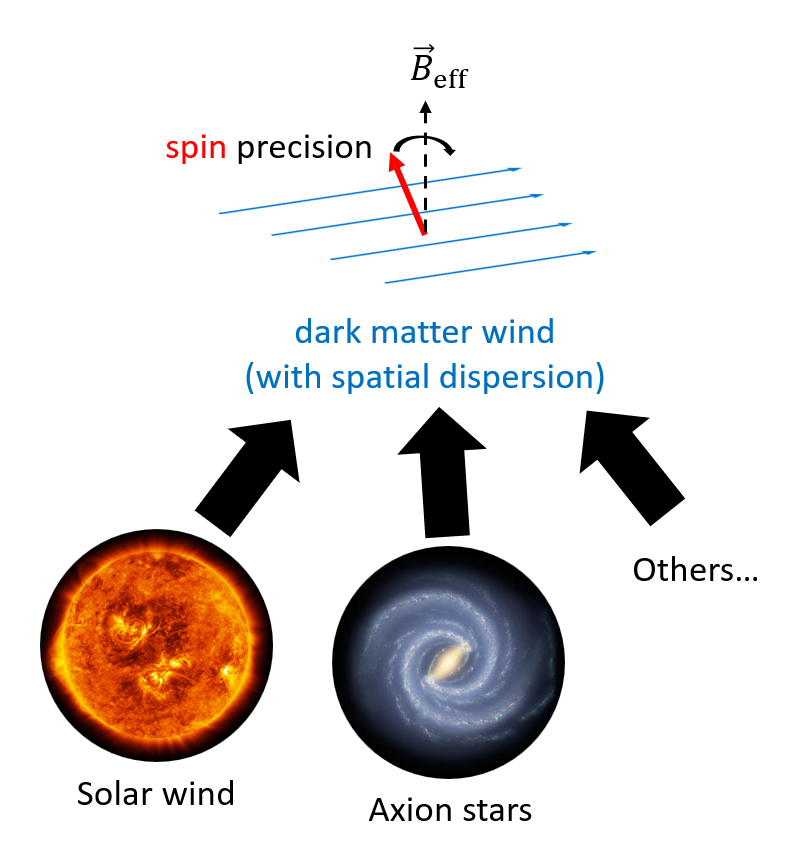}
    \caption{According to Eq.(\ref{eq:ssssss}), the spatial dispersion of the dark matter wind generates an effective magnetic field, which induces a spin precision that can be detected by an atomic magnetometer. The spatial dispersion in the dark matter can be generated from substructures like solar wind or axion stars.}
    \label{fig:graph}
\end{figure}

Although the Earth's spin and the laboratory-made rotation can present a large angular velocity, such treatments only produce an auxiliary rotation on the electrons but not the axions. Since our potential is only valid in inertial frames, such electron rotations contribute nothing to the effective magnetic field. The dark halo model presents the galaxy-scale motion of the dark matter. If the dark matter has some stellar-scale local structures, its angular velocity or local density may be strongly enhanced to give a more accessible signal. If the local dark matter is unevenly distributed, the dispersion of dark matter density will also contribute to the effective angular velocity. Denote the density and velocity profile of dark matter as $f(\mathbf r)$ and $\mathbf v(\mathbf r)$ in the spherical coordinate whose origin is the magnetometer in the laboratory. According to Eq.(\ref{eq:11111}), generally the effective angular velocity should be extracted via integrate
\begin{equation}
    I = \int \mathrm d^3 \mathbf r \ f(\mathbf r)
    \left\{{\boldsymbol{\sigma}} \cdot\left[\mathbf{v}(\mathbf r) \times \hat{\mathbf{r}}\right]\right\}
    \left(\frac{1}{r^2}+\frac{1}{\lambda r}\right) \e^{-r/\lambda}.
\end{equation}
Define the averaged axion density as 
\begin{equation}
    \bar f = \frac{1}{12\pi \lambda^2}\int \mathrm d^3 \mathbf r f(\mathbf r) 
    \left(\frac{1}{r} +\frac 1 \lambda\right)
    e^{-r/\lambda},
\end{equation}
where the preface coefficient ensures that the averaged density equals to itself for a uniform distribution. A factor of $1/r$ inside the integral is extracted into the velocity so that for a uniform angular velocity, the radial integral is completely included in the averaged density. Alternate definitions for the averaged density inside the force range are also applicable according to the need of a specific question. Therefore the effective angular velocity for the effective magnetic field reads
\begin{equation}
    \Omega_{\mathrm{eff}} = \frac{I}{8\pi \lambda^2 \bar f},
\end{equation}
where the coefficient is also determined in the way that for a uniform angular velocity, the effective angular velocity equals itself. It can be immediately seen that not only a non-evenly distributed velocity but also a non-evenly distributed density can generate the effective angular velocity. All the gradients of density and velocity are reflected in the quantity $I$ which is zero for a uniform density and velocity distribution.

For example, a spherically symmetric solution of axion dark matter will present a radius-dependent distribution function $f(r_\s)$ which can be approximately formulated as $f(r_\s)  = f_0 (r_\s/R)^{-\beta}$ at its outer edge. Here, $r_\s$ is the distance from the measuring point to the center of the structure, $R$ denotes the radial scale of the dark matter structure's edge, and $\beta$ depicts how steep its edge is. Realistic examples of dark matter structures following this asymptotic form will be discussed later. Now, we investigate the signal such a structure will generate when it passes through the Earth. For all kinds of dark matter structures and mini-structures theoretically predicted so far, its characteristic scale is at least of the orders of the Earth's radius \cite{banerjee_searching_2020}, which is substantially larger than the interaction range. Within the interaction range $\lambda$, the density distribution can be approximated as an averaged distribution $\bar f$ with a constant gradient $\Delta f = -\bar f \beta /r_\s$ along the radial direction of the structure which can be viewed as a fixed direction within the force range. Assuming that the structure has a fixed velocity $v_s$ inside the force range, after integration using previous formulas, it gives rise to an effective angular momentum of $\Omega_{\mathrm{eff}} = \beta v\sin \xi\sin \zeta\sin\chi/4r_\s$, where $\xi,\zeta,\chi$ are the angles between $\boldsymbol \sigma$ and $\mathbf r_\s$, $\boldsymbol \sigma$ and $\mathbf v$, and $\boldsymbol \sigma \times\mathbf r_\s$ and $\boldsymbol \sigma \times \mathbf v$ respectively, with a positive values for right-hand directions. This particular result shows how a deviation away from the uniform distribution gives rise to a non-zero signal in our model, which is beyond the detection capability of other methods. It also suggests that such a signal has a strong directional dependence. Firstly, it suggests that the signal is strongest when the local velocity is perpendicular to the density gradient. Such a requirement would be automatically satisfied if the structure itself processes a spin and we are sitting at its equator. Otherwise, if the structure has an overall velocity, we need it to skim over the Earth to generate a maximized signal. The signal is also dependent on the direction of the magnetometer, which needs to be aligned along the direction of $\mathbf r_s \times \mathbf v$ to maximize the effective magnetic field signal. More generically, if an inhomogeneous density and velocity are presented simultaneously, we can simply estimate the effective angular velocity as $\Omega_{\mathrm{eff}} \approx \Delta(nv)/n\lambda$ where $\Delta(nv)$ is the number density and velocity difference within the force range $\lambda$. The angular dependence of such complicated signals needs meticulous investigation for specific problems.

There have been various theoretic models that predict different structures of the dark matter that potentially exhibit such effective angular momentum signals as discussed previously. Such local structures can be generated thorugh interacting with celestrial bodies constructed by standard model particles, such as interacting with the Sun or Earth via scattering or gravitation. When dark matter travels through the Sun, it will be scattered by numerous leptons or baryons in the Sun. As depicted in this article, since standard model particles like electrons experience forces from dark matter axions, axions will also be significantly influenced or emitted by dense matter. The dark matter flow generated by such progress is called the dark solar wind and might be strong enough to be detected \cite{an_solar_2021,redondo_solar_2013}. The number density of the dark solar wind is predicted to be around three orders higher than the background dark matter \cite{chang_dark_2022}. To generate a detectable angular velocity, it requires the velocity of dark matter wind to have a gradient of about $10^{-7}\ \mathrm{m/s}$ inside the $1\ \mathrm{m}$ force range, coupled with the amplification of number density. Given that the typical velocity of non-perturbed dark matter is around $v\approx 200\ \mathrm{km/s}$, such a velocity gradient might possibly be realistic. Gravitational potential wells of the Earth and the Sun can attract dark matter axions and form a dark halo around them \cite{banerjee_searching_2020}, similar to the formation mechanism of galactic dark halos. Such halos can also provide a larger number density than in non-perturbed cases. Since the dark halo is locally bounded, its density will decrease radially and have a radial gradient around $\sim 1/R$, where $R$ denotes the radius of the Earth or the Sun. As we have calculated previously, such a density gradient will generate an effective angular velocity of $v/R$, where $v$ is the local velocity of the dark matter halo. As modeled in \cite{banerjee_searching_2020}, if the dark halo's rotation axis is aligned with the polar axis of the Earth, we would expect a maximum signal of approximately $\Omega_{\mathrm{EarthHalo}}\approx \alpha \Omega_{\mathrm{E}}$ near the equator, where $\Omega_{\mathrm{E}} \approx 7 \times 10^{-5}\ \mathrm{s^{-1}}$ is the rotational angular velocity of the Earth and $\alpha$ is the amplification of the Earth's dark halo density with respect to the background dark matter. 

Axion stars formed by self-interactions and gravitational forces between axions are also proposed as a possible substructure of the dark matter \cite{braaten_colloquium_2019,arvanitaki_large-misalignment_2020}. Apart from the mass term in the equation of motion of axions, there would be a four-particle self-interaction term as a first-order approximation \cite{Marsh_axion_2016}, which exhibits an additional attracting force. Together with the gravitational force, the axions are attracted by each other and form celestial bodies at the scale of galaxies or stellar systems \cite{jens_c_niemeyer_small-scale_2020}. These kinds of structures can be roughly understood as soliton-like stationary solutions of its equations of motion, with an almost flat central core and a steep outer gradient. There could be various possible stellar-scale structures like clumps or streams in the dark matter \cite{diemand_clumps_2008}. Other theories state that highly compact axion miniclusters can form out of large isocurvature perturbations after the end of the cosmological QCD phase transition \cite{zurek_astrophysical_2007,hardy_miniclusters_2017,jens_c_niemeyer_small-scale_2020}. The outer regions of these possible ministructures have a steep slope whose density profile can be generally fitted as $f(r_\s) = f_0 (r_\s/R)^{-\beta}$ as we used in the previous calculations \cite{jens_c_niemeyer_small-scale_2020}. Some recent studies investigate the formation of such structures in depth using N-body simulations, giving a density profile with an exponent $\beta \approx 1$ \cite{delos_are_2018,delos_density_2018}. We note that these theories and predictions of axion ministructures are highly model- and parameter-dependent, hence giving a wide range of possible effective angular momentum for detection. Nevertheless, they still provide a solid basis for detection by experiments.

Our method hence provides a new portal to detect the substructures in the dark matter as illustrated in FIG.\ref{fig:graph}. For experimental implementation, an alignment of atomic magnetometers covering different places around the Earth can be established, as the experimental setup in the Global Network of Optical Magnetometers for Exotic physics searches (GNOME) \cite{afach_search_2021,afachWhatCanGNOME2024}. Such a setup has the ability to discover the angular- and position-dependence of the dark matter substructures, especially for detecting the Earth's halo. By deploying these magnetometers into a wider range, like solar orbits, the detection of larger structures like the solar halo would be possible. We can also search for dark matter structures by tracking the movement of such background signals on the alignment magnetometers. Due to the position-dependence of the ministructure signals, as we discussed before, such a widely-ranged alignment can increase the chance of finding a non-zero signal and excluding false signals. On the other hand, to detect possible dark matter structures generated by the Sun or other heavy celestial bodies, we can deploy a detector near these celestial bodies and search for possible signals stimulated by them. Their perturbations to dark matter would be numbers of orders stronger than the Earth, hence providing a greater chance of discovering new physics.

\section{Conclusion}
In this work, we introduce a new proposal that utilizes the exotic interaction between dark matter axions and standard model fermions to investigate the potential makeup and local structures of dark matter. It can also help to constrain the parameters of new particles beyond the standard model. We discuss the example of the pseudoscalar axion dark matter model, highlighting the presence of long-range magnetic-field-like interactions between dark matter axions and electrons or nucleons. This interaction can be seen as an effective magnetic field acting on polarized electrons or nucleons, as shown in Eq.(\ref{eq:ssssss}). Consequently, within this model, it becomes theoretically feasible to detect a non-zero magnetic field signal using atomic magnetometers designed based on electron or nuclear spin. This method can provide estimates that are close to or even surpass the upper limits set by axion-mediated fermion-fermion long-range interactions. Particularly within the low axion mass range, as our method is more sensitive to the axion mass and is not confined by the crystal volume employed in the experiment, it can achieve an upper limit that is two orders of magnitude higher than previous results. With the development of magnetometers, our theory can be applied to a wider range of axion masses, with an upper bound on the coupling coefficient exceeding the astrophysics limits. Furthermore, since the strength of this effective magnetic field is contingent on the angular velocity of dark matter, it becomes feasible to measure the rotation and local structures of dark matter. Differing from previous methods that mainly focus on the universe-scale and galaxy-scale motion of dark matter, our method provides an efficient way to investigate the stellar-scale and laboratory-scale mini-structures of dark matter. The effective angular velocity of dark matter is highly model-dependent, which needs further investigation.

The method discussed in this article also provides inspiration for future research. Firstly, the interaction we calculated is Lorentz covariant, meaning that this potential does not apply in a rotating frame. If we want to separate the angular velocity of dark matter and derive an independent upper limit measurement of the coupling coefficient, one approach is to introduce an artificial rotation. In this case, the interaction form needs to be calculated based on field theory in a rotating frame. Secondly, similar to the work in Ref. \cite{dobrescu_spin-dependent_2006}, this interaction can be applied to other dark matter models and other new particles beyond the standard model. Furthermore, compared with detecting dark matter through spin- or velocity-independent scattering cross sections, our model can yield accurate results with the high sensitivity of atomic magnetometers. It is also more sensitive to the specific composition of dark matter and capable of distinguishing the types of dark matter particles.

\appendix
\section{Conventions}
In our derivation, the Dirac matrices are defined using a chiral basis as follows:
\begin{equation}
    \gamma^\mu = \left(
    \begin{array}{cc}
        0 & \sigma^\mu \\
        \bar\sigma^\mu & 0
    \end{array}
    \right),\quad \mu = 0,1,2,3,
\end{equation}
and
\begin{equation}
    \gamma^5 = \ii \gamma^0\gamma^1\gamma^2\gamma^3= \left(
    \begin{array}{cc}
        - 1_{2\times 2} & 0 \\
        0 & 1_{2\times 2}
    \end{array}
    \right),
\end{equation}
where 
\begin{equation}
    \sigma^\mu = (1_{2\times 2},\hat{\boldsymbol \sigma}),\quad \bar \sigma^\mu = (1_{2\times 2},-\hat{\boldsymbol \sigma})
\end{equation}
is the four-vector form of the Pauli matrices, $1_{2\times 2}$ denotes the $2\times 2$ identity matrix, and $\hat{\boldsymbol\sigma}$ denotes the three-component Pauli matrices. Here we put a hat explicitly to distinguish with the spin vector $\boldsymbol{\sigma}$. The Pauli matrices are defined on the standard basis as 
\begin{equation}
    \begin{aligned}
        \hat\sigma^1 = \left(\begin{array}{cc}
            0 & 1 \\
            1 & 0
        \end{array}\right),\ 
        \hat\sigma^2 = \left(\begin{array}{cc}
            0 & -\ii \\
            \ii & 0
        \end{array}\right),\ 
        \hat\sigma^3 = \left(\begin{array}{cc}
            1 & 0 \\
            0 & -1
        \end{array}\right).
    \end{aligned}
\end{equation}
In this basis, the particles' normalized Dirac spinors can be written
\begin{equation}
        u(p) = \left(
        \begin{array}{c}
             \sqrt{p\cdot\sigma} \xi\\
             \sqrt{p\cdot \bar\sigma}\xi
        \end{array}
        \right),
\end{equation}
where $\xi$ is the two-component spinor normalized to unity. We have, for example, $\xi = \left(\begin{array}{cc}
     1 \\
     0 
\end{array}\right)$ for spin up in the $z$ direction and $\xi = \left(\begin{array}{cc}
     0 \\
     1 
\end{array}\right)$ for spin down in the $z$ direction.

\section{Derivation of spinor products}
The scattering amplitude in Eq.(\ref{eq:111}) can be decomposed into summations of spinor products presented in Eq.(\ref{eq:sppr1}) and Eq.(\ref{eq:sppr2}). Since the spatial part of the propagator's momentum $l$ equals zero, there's no contribution from products of the form $\bar u(p')\gamma^j u(p)$ where $j=1,2,3$. From our conventions of the spinors in Appendix A, the spinor products in Eq.(\ref{eq:sppr1}) and Eq.(\ref{eq:sppr2}) give
\begin{equation}
    \bar u(p') u(p) = \xi^\dagger \left(
        \sqrt{p'\cdot \bar \sigma p\cdot \sigma} + \sqrt{p'\cdot \sigma p\cdot \bar\sigma}
    \right)\xi,
    \label{eq:uu}
\end{equation}
\begin{equation}
    \bar u(p') \gamma^0 u(p) = \xi^\dagger \left(
        \sqrt{p'\cdot \bar \sigma p\cdot \bar\sigma} + \sqrt{p'\cdot \sigma p\cdot \sigma}
    \right)\xi,
    \label{eq:u0u}
\end{equation}
where $\xi$ is the two-component spinor of the electron. Since the axion is spinless, the scattering process will not alter the electron's spin. Using the definitions of $P$ and $q$ in the main article, the first square root in Eq.(\ref{eq:uu}) reads
\begin{equation}
    p'\cdot \bar \sigma p\cdot \sigma = E_\e^2  - \left(\mathbf P^2-\frac{\mathbf q^2}{4}\right) + \left[
        \mathbf q - \ii \left(
            \mathbf q\times \mathbf P
        \right)
    \right]\cdot \hat{\boldsymbol \sigma},
\end{equation}
and other functions inside the square roots can be derived similarly. The square root of a general $2\times 2$ Hermitian matrix reads
\begin{equation}
    \sqrt{c + \mathbf d\cdot \hat{\boldsymbol\sigma}} = 
    \sqrt{b} (1 + \mathbf a\cdot \hat{\boldsymbol{\sigma}}),
    \label{eq:rtsigma}
\end{equation}
where 
\begin{equation}
    b = \frac{c+ \sqrt{c^2-\mathbf d^2}}{2},
\end{equation}
and
\begin{equation}
    \mathbf a = \frac{\mathbf d}{2b}.
\end{equation}
These formulas can be easily checked by squaring the right-hand side of Eq(\ref{eq:rtsigma}). The square roots in Eq.(\ref{eq:uu}) can thus be calculated by these formulas, and the equation is reduced to
\begin{equation}
    \bar u(p')u(p) \approx 2\sqrt{E_\e^2-\mathbf P^2}\left(
        1-\frac{\mathrm i }{2(E_\e^2-\mathbf P^2)}(\mathbf q\times \mathbf P)\cdot(\xi^\dagger\hat{\boldsymbol \sigma} \xi)
        \right),
\end{equation}
where we keep the lowest order terms for its scalar part and spin-dependent part separately. The Pauli matrices $\hat{\boldsymbol \sigma}$ sandwiched between spinors $\xi$ gives the expectation value of the electron's spin polarization $\xi^\dagger\hat{\boldsymbol \sigma} \xi = \boldsymbol{\sigma}$. The spinor product with $\gamma^0$ in Eq.(\ref{eq:u0u}) is calculated in the same manner and its result is shown in Eq.(\ref{eq:sppr1}).


%

\end{document}